\documentclass[]{spie}  
\usepackage[dvips]{graphicx}

\title{The distribution of mass ratios in compact object binaries} 

\author{Tomasz Bulik\supit{a}, Krzysztof Belczynski\supit{b,c}
 and Vassiliki Kalogera\supit{b}
\skiplinehalf
\supit{a}Nicolaus Copernicus Astronomical Center, Bartycka 18, 00176 Warsaw, Poland\\
\supit{b} Northwestern University, 
       2145 Sheridan Rd., Evanston, IL 60208, USA\\
\supit{c}Lindheimer Fellow
}

\authorinfo{E-mail:
T. B.:  bulik@camk.edu.pl;
K. B.: kbelczyn@northwestern.edu;
V. K.: vicky@northwestern.edu}

\begin{document} 
  \maketitle 

\begin{abstract}

Using the StarTrack population synthesis code we compute the distribution
of masses of merging compact object (black hole or neutron star) binaries.
The shape of the mass distribution is sensitive to some of the parameters
governing the stellar binary evolution. We discuss the possibility of
constraining stellar evolution models using mass measurements obtained 
from the detection of compact object inspiral with the upcoming 
gravitational-wave observatories. 

\end{abstract}



\section{INTRODUCTION}

The  operational phase of the  gravitational
wave detectors  LIGO\cite{1992Sci...256..325A} and VIRGO \cite{1990brada}
is quickly approaching, and 
in a few years we anticipate the operation  of the more
sensitive advanced LIGO detector. Mergers of compact objects
in binaries are the most promising sources of gravitational waves
to be detected by these detectors. The 
expected rates of such mergers have been 
a subject of intense research. 
Recent  analyses \cite{2001ApJ...556..340K,2002ApJ...572..407B,KKL}
have shown there is a little chance  that 
 LIGO should see any mergers of compact object binaries, while LIGO~II
could detect a substantial number, up to hundreds or thousands of such
events. These estimated rates however, do vary strongly with the model of
stellar binary evolution.

Once we begin to detect such events, the data analysis should yield
information about each system that merged. The merging of compact object
binaries proceeds in three phases: inspiral, merger, and ringdown. It is
the inspiral phase that will be used for detection of the stellar masses
involved.  The analysis of the frequency change with time of the event
should yield the masses of the merging objects\cite{will}. However,
determining their type, i.e. a neutron star or a black hole will be
difficult\cite{blanchet} with the measurement during this phase. The
modifications of the gravitational wave signal due to the mass
distribution of the neutron star are small and may amount to one
 in $10^4$ oscillations.

The analysis of the observation of merger phase may lead to determination
of the type of the compact object i.e. will be different for mergers
involving neutron star, and may even lead to constraints on the neutron
star equation of state. Also, if the neutron stars are rotating there is a
chance of exciting stellar oscillations by the orbital motion which may
modify the gravitational waveform\cite{1999MNRAS.308..153H}.

It is then important to ask what is the expected two dimensional
distribution of masses of compact object binaries that merge, and how does
this distribution depend on the assumed model of stellar binary evolution.
In \S\,2 we present the dependence of the mass ratio distribution on the
particular stellar evolution model. Next, \S\,3 is devoted to the
potential observations. The main observable derived from the
gravitational-wave signal is the chirp mass. We show the dependence of the
distribution of the observed chirp masses on the stellar evolution
parameters and estimate the possibility of distinguishing between the
models using the amount of data that we expect to be obtained. Finally in
\S\,4 we summarize our results.

\begin{figure}[t]
\centerline{
\includegraphics[width=0.9\textwidth]{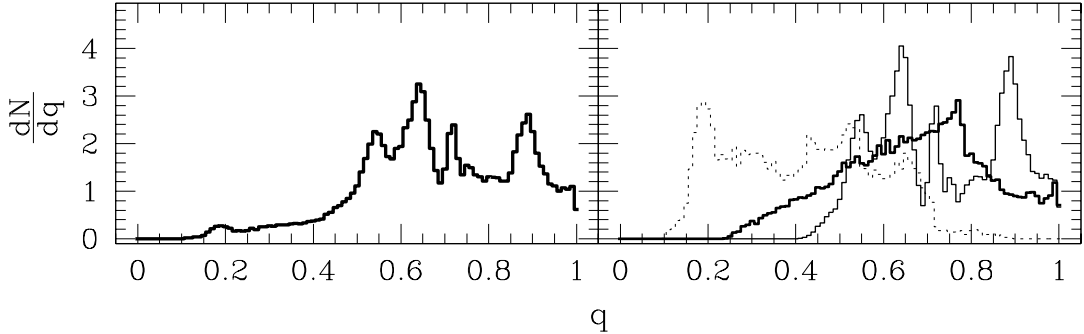}}
\caption{The distributions of intrinsic mass ratio in compact objects
produced in the standard model simulation (left panel), and
the contributions of different types of binaries (BH-BH thick line, NS-NS
thin line , and 
BH-NS dotted line)to this
distribution. The distributions  have been 
calculated using a total of 106307 compact object binaries
and the  bin  width is $dq=0.01$.
  Each  distribution is normalized to unity. 
 }
\label{rysone}
\end{figure}

\section{The distribution of mass ratio in compact object binaries.}

In order to calculate theoretical distributions of the mass ratio of
compact object merging events we use the StarTrack population synthesis
code \cite{2002ApJ...572..407B}. The standard evolutionary scenario of the
code uses modified formulae of Ref.~\citenum{2000MNRAS.315..543H} to
evolve single stars along with their proposed wind mass loss rates. The
code takes into account different types of binary interactions:  
dynamically stable and unstable mass transfer phases. The dynamically
stable events are modeled using the formalism of
Ref.~\citenum{1992ApJ...391..246P}, while unstable mass transfers are
treated following Ref.~\citenum{1984ApJ...277..355W}.  Compact object
(black holes, neutron stars) formation in supernova (SN) explosion/core
collapse events, is described in the code following the results of
hydrodynamical calculations of Ref.~\citenum{1999ApJ...522..413F}. Neutron
stars receive natal kicks from the distribution of
Ref.~\citenum{1998ApJ...505..315C}; intermediate mass black holes are
formed through partial fall back and in attenuated SN explosions receiving
smaller kicks than neutron stars; while most massive black holes are
formed through direct collapse of their immediate progenitors with no
accompanying SN explosion and they do not receive any kicks at their
formation.
 
The distribution  of masses in compact object 
binaries can be described by the probability density 
$\rho ( m_1, m_2)$.
The population synthesis code provides  a numerical
representation of this distribution 
 \begin{equation}
 \rho( m_1, m_2) = {1\over N} \sum_i \delta(m_1^i)\delta(m_2^i)
 \label{rho}
 \end{equation}
where $\delta$ is the Dirac's function, 
$N$ is the number of binaries obtained in a simulation.

 We present the mass ratios distributions
\begin{equation}
 {dp\over dq} = \int dm_1\int dm_2    {\rho( m_1, m_2)} 
  \delta\left(q-{m_1\over m_2}\right) = {1\over N} \sum_i\delta(q_i)
\end{equation}  
 of different
types of compact object binaries in the right panel
of Figure~\ref{rysone}, and in the left panel we show the
total distribution summed over all the objects. Note that
the total distribution is  dominated by
double NS binaries. They 
are more numerous than the binaries containing black
holes simply because of the steep fall of the initial mass
function\cite{2002ApJ...567L..63B}.
The distributions are not flat and exhibit 
distinctive features which are not numerical artifacts; the
statistical 
fluctuation in the left panel of Figure~\ref{rysone} amount to less than 1\%.

The StarTrack code allows to test different 
models of stellar binary evolution. We have run a set of simulations with
different models  to show the dependence
of the observed distribution of mass ratio
on a given evolutionary model. We list the 
models and their description in Table \ref{tone}.

\begin{table}
\caption{Description of different population synthesis models 
for which the distributions of mass ratios have been calculated. For more 
details see Ref.\ 3.}
\label{tone}
\begin{center}
\begin{tabular}{ll}
\hline \hline
Model & Description \\
\hline
 
A      & standard model described in  Ref. 2. \\
B1,4,6,9 & zero kicks, single Maxwellian with \\
       & $\sigma=30,50,300,$\,km\,s$^{-1}$, \\
       
C      & no hyper--critical accretion onto NS/BH in CEs \\
D1--2  & maximum NS mass: $M_{\rm max,NS}=2, 1.5$\,M$_\odot$ \\
E1--3  & $\alpha_{\rm CE}\times\lambda = 0.1, 0.5, 2$ \\
F1--2  & mass fraction accreted: f$_{\rm a}=0.1, 1$ \\
G1--2  & wind changed by\ $f_{\rm wind}=0.5, 2$ \\
H1--2     & Convective Helium giants: $M_{\rm conv}=4.0$,\, $0.0\,M_\odot$ \\
I      & burst--like star formation history \\
J      & primary mass: $\propto M_1^{-2.35}$ \\
L1--2  & angular momentum of material lost in MT: $j=0.5, 2.0$\\
M1--2  & initial mass ratio distribution: $\Phi(q) \propto q^{-2.7}, q^{3}$\\
N      & no helium giant radial evolution\\
O      & partial fall back for $5.0 < M_{\rm CO} < 14.0 \,M_\odot$\\
\hline
\end{tabular}

\end{center}
\end{table}

\begin{figure}[t]
\centerline{
\includegraphics[width=0.9\textwidth]{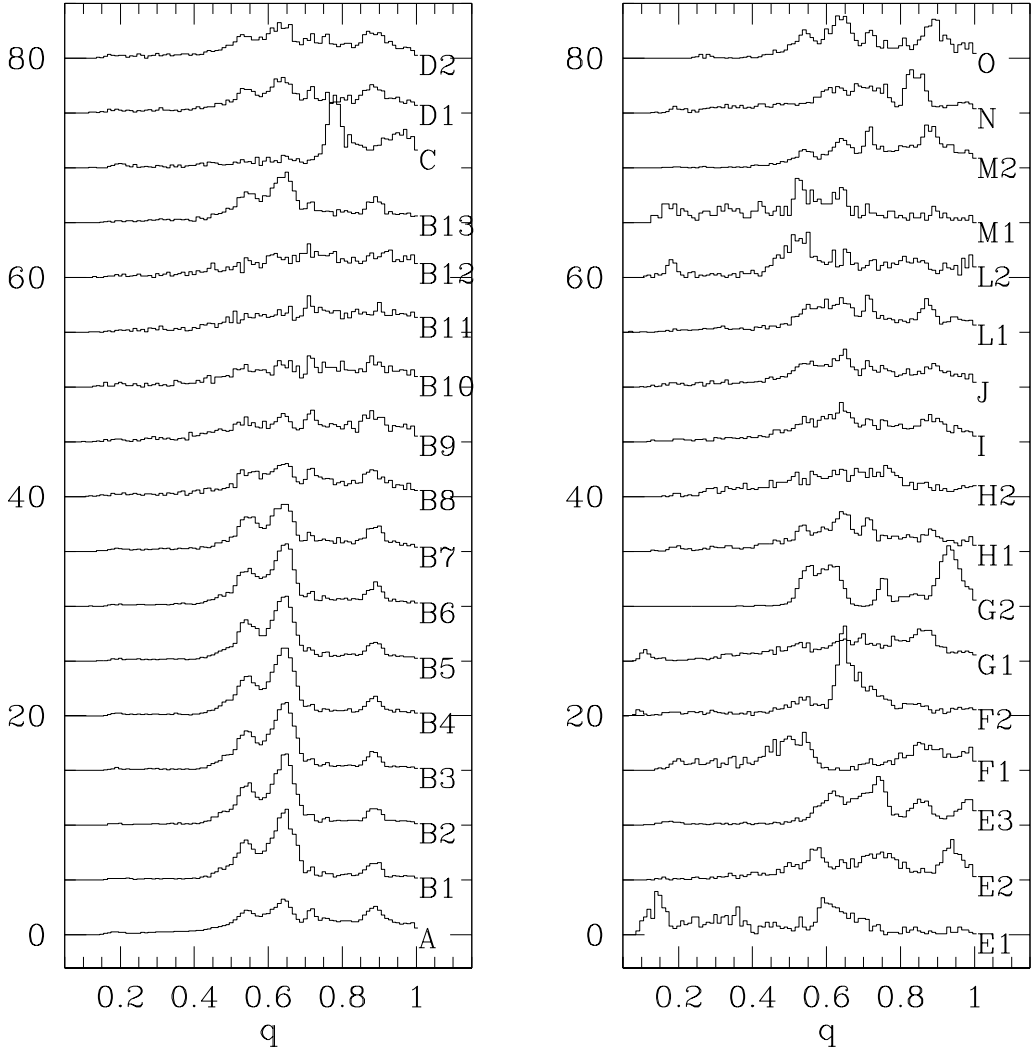}}
\caption{
Distributions of mass ratio for a set of models listed in 
 Table~\ref{tone}. The distributions are  normalized to unity, but for clarity
  we added $5$ to each consecutive distribution. 
 }
\label{rystwo}
\end{figure}

We present the distributions of the mass ratios for a range of stellar
evolutionary models in Figure~\ref{rystwo}. The distributions vary
significantly with the particular choice of the model. The kick velocities
(models B) change the relative strength of the peaks in the distribution.
These peaks are related to the neutron star in binaries, see
Figure~\ref{rysone}.  Various parameters lead to specific changes in the
distribution. IN particular several parameters influence the shape of the
distribution very significantly: variation of the common envelope
efficiency $\alpha_{CE}\lambda$ (models E), changes of the stellar wind
strength (models G) etc.  The distribution of masses, and consequently
mass ratios of compact object binaries does carry significant information
about the evolution of the massive binaries. It is therefore important to
ask the question: Can we constrain observationally this distribution? Is
it possible to learn about the evolution of massive stars by measuring
masses of merging binaries?

\section{What can be observed by GW detectors? }

Mergers of compact object binaries will be observed 
by the gravitational wave interferometers during 
the inspiral phase, just  minutes prior to the merger.
The data train will  allow us to measure at least the chirp mass:
\begin{equation} 
{\cal M}(m_1,m_2) = (m_1 +m_2) \left( {m_1\over m_2} +{m_2\over m_1}\right)^{-3/5} =
m_2 (1 +q) \left( {q +q^{-1}}\right)^{-3/5}
\end{equation}
The individual masses of merging compact objects, $m_1$ and $m_2$, 
 will be measured if higher
order correction due to general relativity are taken into account. 
To find the 
distribution of the observed chirp masses we need to 
take into account also the different strength of the gravitational
wave signal from sources with different mass.

The strength of the gravitational wave signal 
from coalescing compact sources has been estimated in 
Refs~\citenum{1993PhRvD..47.2198F,1998PhRvD..57.4535F}. In the inspiral phase 
the  signal to noise $S/N$ ratio 
for LIGO interferometers is proportional to ${\cal M}^{5/6}$.
In particular:
\begin{equation}
\left( {S\over N}\right) = k \left( {1 {\rm Gpc} \over D}\right) 
\left( {4\mu \over M}\right)^{1/2}
\left( {(1+z) M\over 18 M_\odot}\right)^{5/6}
\label{SN}
\end{equation}
where $D$ is the distance, $M$ is the total mass 
of the system, $\mu$ is the reduced mass, 
 $k= 0.58$ for the initial LIGO, and $k=15.3 $ for the advanced LIGO
configuration. Therefore the sampling distance for a constant 
signal to noise ratio is proportional to 
$ D\propto \left( {4\mu / M}\right)^{1/2} M^{5/6}$, and the sampling
volume $V\propto D^3$.
Thus the observed number of sources observed 
with given masses $m_1$ and $m_2$ is
\begin{equation} 
dN= V(m_1,m_2) \rho(m_1,m_2) dm_1 dm_2\, .
\end{equation}
and the total number of sources to be observed is
$N_{tot}=\int dN$. The normalized distribution of the observed
chirp masses  is
\begin{equation}
p({\cal M})={1\over N_{tot}}{dN\over d{\cal M}} 
={1\over N_{tot}}
\int dm_1 \int  dm_2  V(m_1,m_2) \rho(m_1,m_2) 
\delta\left({\cal M}- {\cal M}(m_1,m_2)\right)\,.
\label{pq}
\end{equation}
The population synthesis code provides us with a numerical 
representation of the $\rho(m_1,m_2)$ distribution from which we obtain the
numerical representation of the  distribution of the mass
ratios $p({\cal M})$, 
by inserting Equation~\ref{rho} into Equation \ref{pq}, and 
replacing integrals with sums.

\begin{figure}
\includegraphics[width=0.9\textwidth]{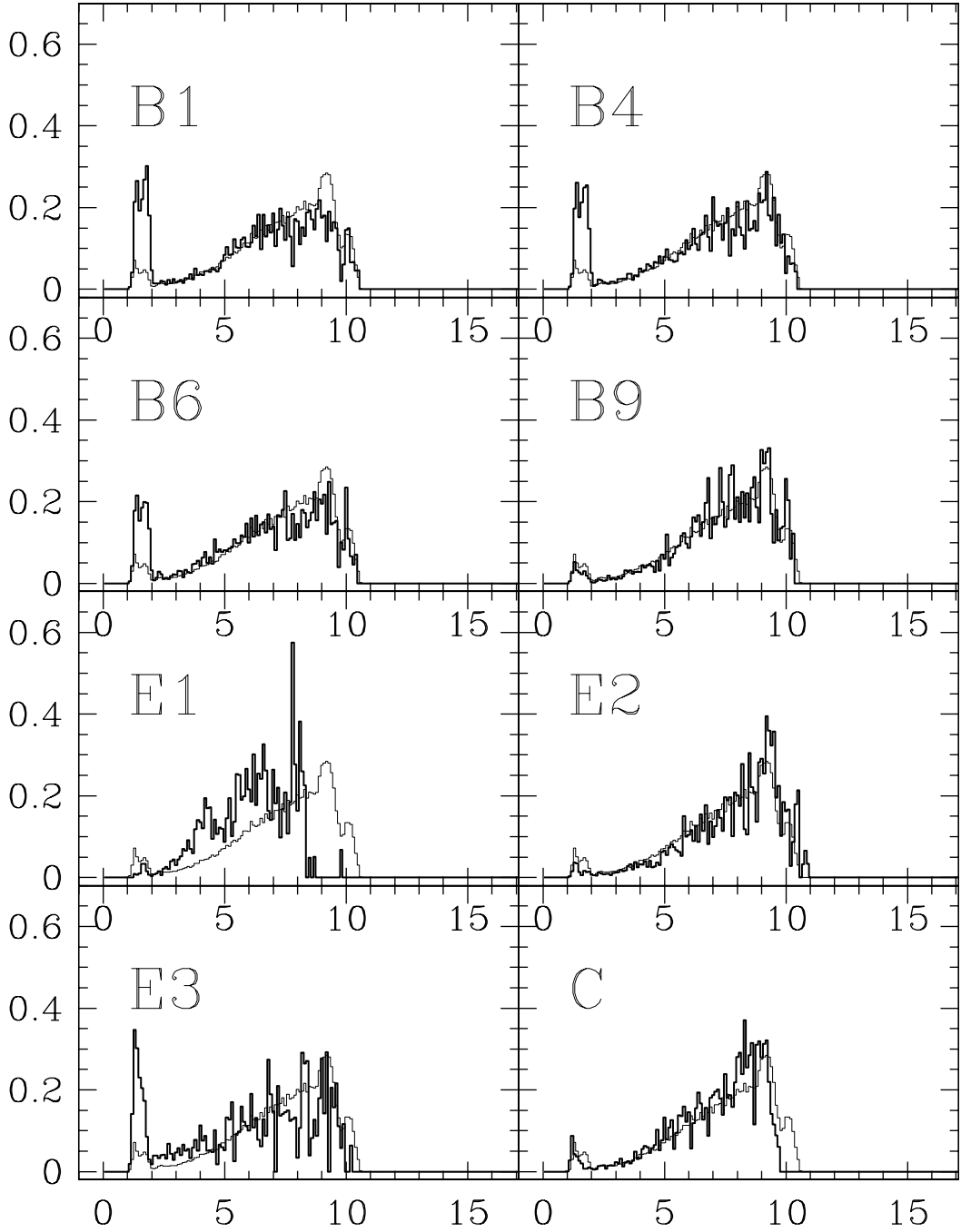}
\caption{The expected distributions of the observed chirp mass 
in selected models with varying kick velocity distributions (B1,B4,B6,B9), with
different common envelope efficiency $\alpha_{CE}\lambda$ (E1, E2, E3)
and without the hypercritical accretion onto the compact object (C). In each
panel we present for comparison the distribution of model A .}
\label{one}
\end{figure}

\begin{figure}
\includegraphics[width=0.9\textwidth]{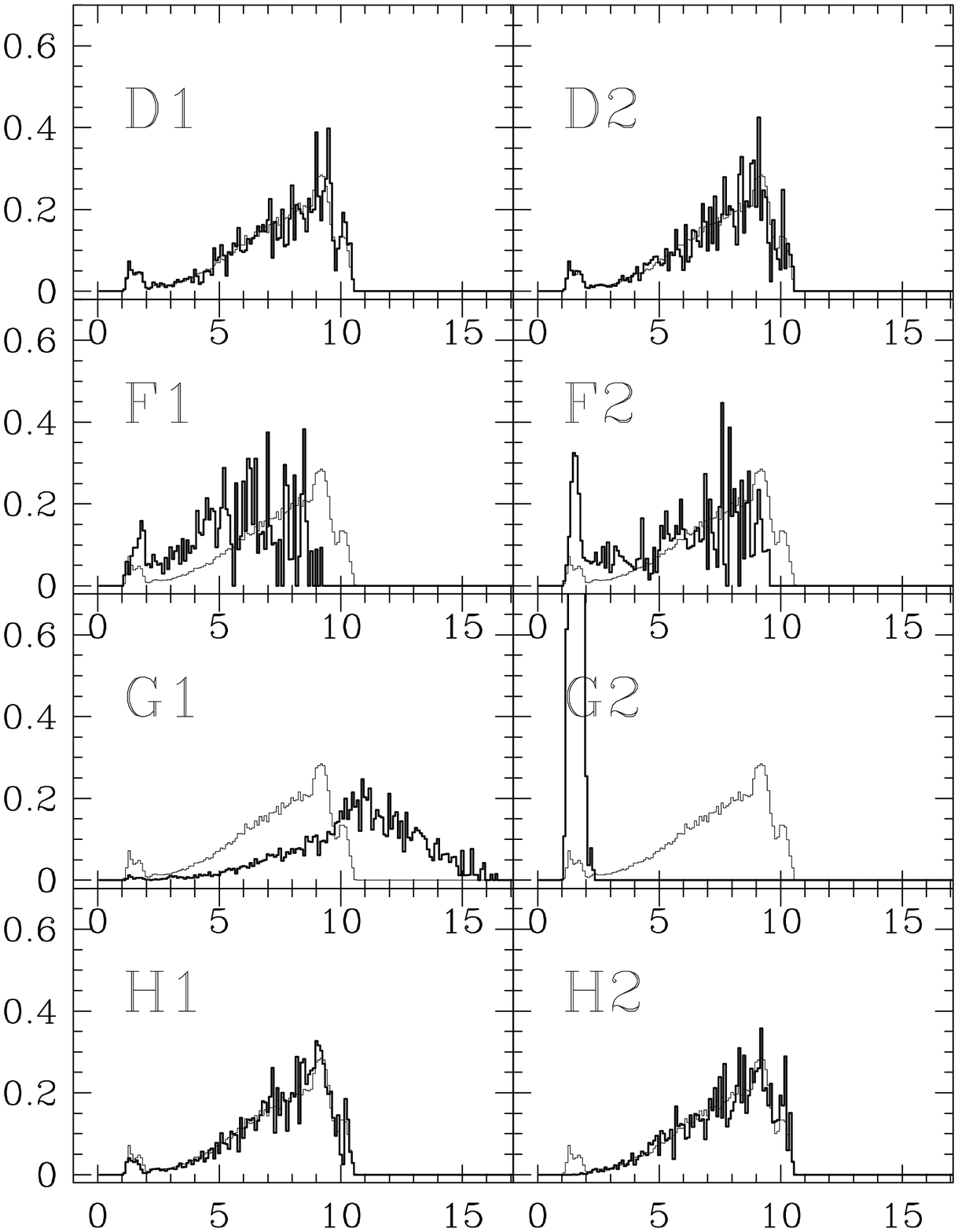}
\caption{The expected distributions of the observed chirp mass 
in models  D,F,G,H. In each
panel we present for comparison the distribution of model A . }
\label{two}
\end{figure}

\begin{figure}
\includegraphics[width=0.9\textwidth]{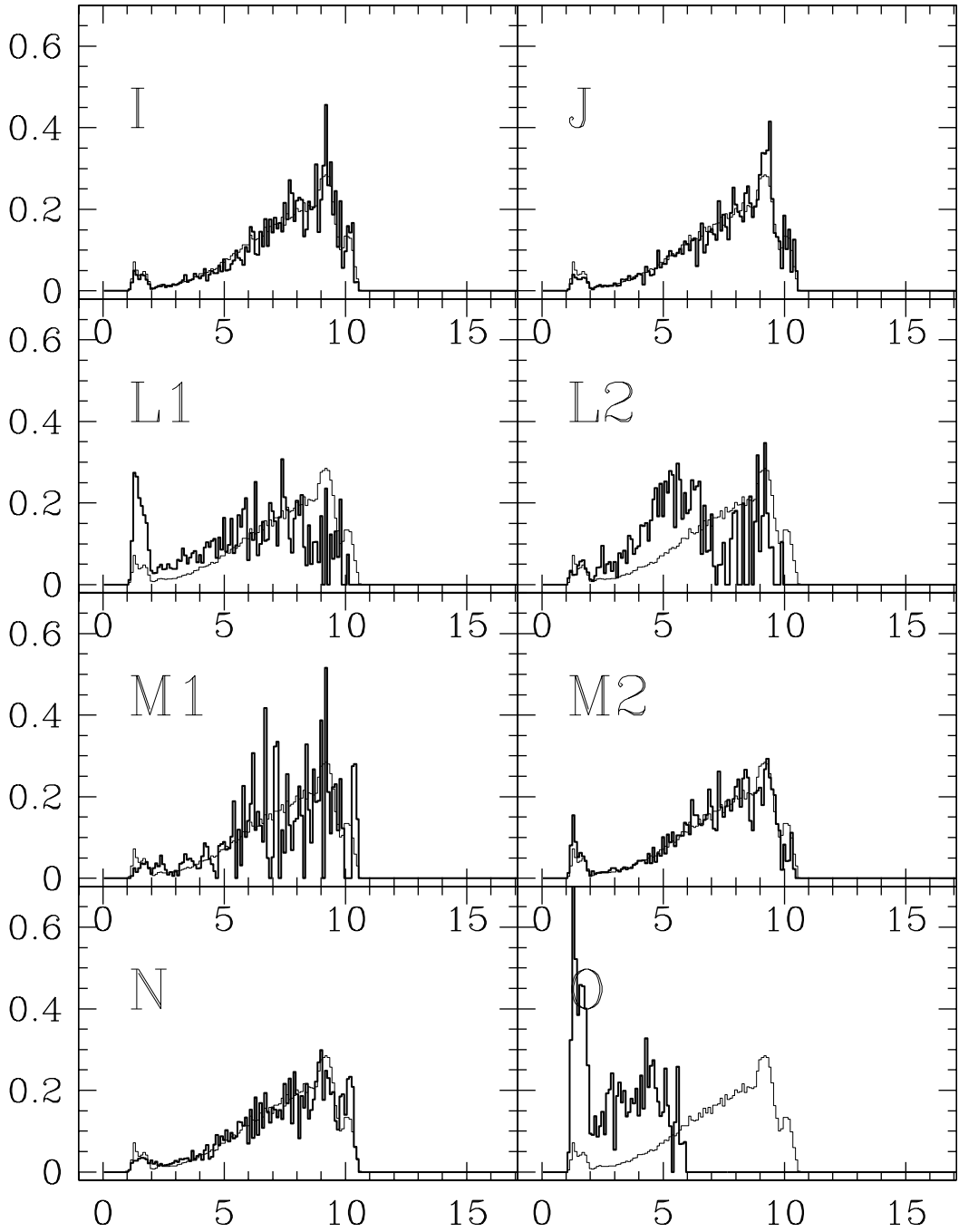}
\caption{The expected distributions of the observed chirp mass 
in models I, L, M, N, O.  In each
panel we present for comparison the distribution of model A .}
\label{three}
\end{figure}

We present the expected distributions of observed chirp masses in
Figures~\ref{one}, \ref{two}, and~\ref{three}. The dependence of the
observed distribution of mass ratios on the kick velocity is presented in
the top panels of Figure~\ref{one}. The observed distribution is dominated
by the BH-BH mergers, and BHs do not receive substantial kicks, especially
for the massive BH systems for which the sampling volume is the largest.
However, the contribution of binaries containing neutron stars at ${\cal
M}\approx 1.5\, M_\odot$ changes their number, see the strength of the
peak at ${\cal M} \approx 1-2\,M_\odot$.

The bottom panels of Figure~\ref{one} show the dependence of the observed
distribution of chirp masses ratios on the common envelope efficiency
parameter $\alpha_{CE}\lambda$.  Turning off the hypercritical accretion
onto compact sources (model C), is shown in the bottom right panel of
Figure~\ref{one}).

Models D1 and D2 shown in the top panels of Figure~\ref{two} serve as self
consistency checks of the calculations. In these models the maximum mass
of a NS is decreased, and since the character of the gravitational wave in
the inspiral phase does not depend on the nature (BH or NS) of a given
compact object they should not differ from the standard model A. In models
F1 and F2, shown also in Figure~\ref{two}, we vary the mass fraction
accreted onto a compact object in mass transfer events. While the
statistic for this models is poor we can clearly say that the observed
chirp mass distributions vary strongly. Models G1 and G2, shown in
Figure~\ref{two} correspond to the increasing and decreasing of the
stellar winds by a factor of two. The decreased wind strength shifts the
distribution to the higher values and decreases the number of small chirp
mass ratio mergers, while increasing the stellar winds makes NS-NS mergers
dominate the observed sample, and the entire distribution shifts to the
lower values. Models H1 and H2, where we test different values of
$M_{conv}$ of helium giants affect only the properties of the neutron
stars and therefore do not strongly change the overall shape of the
distribution, save for the peak at ${\cal M} \approx 1-2\,M_\odot$.

Models L, presented in  Figure~\ref{three} correspond to variation of the 
angular momentum fraction lost in mass transfer events, and  affect strongly
the shape of the observed distribution of chirp masses. Models M, where we vary
the shape of the initial mass ratio  distribution do not make a significant
difference. This is because  the compact object binaries are formed from
binaries which initially had rather well defined properties (like the initial
mass ratio), the number of such binaries does change fore different initial
mass ratio distributions but their properties do not. Model N where we neglect
the helium giant radial evolution affects the properties of neutron stars and
does not  change significantly the distribution of observed chirp masses.
Finally in model O we change the prescription  for masses of the newly formed
compact objects, which  has a strong influence on the shape of the distribution
of observed chirp masses.

\begin{table}[t]

\caption{Results of the  simulations of 
observation of $20$, $100$ and $500$ mergers. We list 
KS-test probabilities that appeared in less than 1\% of
tries. The lower the probability  the easier can a given model 
be distinguished.
}
\begin{center}
\begin{tabular}{lrrr}
Model  &   20 mergers  & 100 mergers  & 500 mergers \\
\hline\hline
 B1  & $    0.943   $ & $   0.169       $ & $   2.19\times 10^{-8} $\\
 B2  & $    0.945    $ & $    0.261     $ & $     3.56\times 10^{-9}$\\
 B3  & $    0.966    $ & $    0.483     $ & $    1.19\times 10^{-6}$\\
 B4  & $    0.956    $ & $    0.523     $ & $    1.11\times 10^{-6}$\\
 B5  & $    0.971   $ & $    0.464      $ & $    4.44\times 10^{-5}$\\
 B6  & $    0.980    $ & $    0.327     $ & $    9.20\times-07$\\
 B7  & $    0.990    $ & $    0.891     $ & $    5.28\times 10^{-2}$\\
 B8  & $    0.991    $ & $    0.977     $ & $   0.625   $\\ 
 B9  & $    0.991    $ & $    0.989     $ & $   0.823   $\\ 
 B10 & $   0.988     $ & $   0.886     $ & $   0.138   $\\ 
 B11  & $   0.984    $ & $    0.580     $ & $    2.45\times 10^{-5}$\\
 B12  & $   0.973    $ & $    0.286     $ & $    1.77\times 10^{-6}$\\
 B13  & $   0.995    $ & $    0.990     $ & $   0.870   $\\
 C  & $     0.966    $ & $    0.135     $ & $   1.48\times 10^{-6}$\\
 D1  & $    0.992    $ & $    0.993     $ & $  0.750    $\\
 D2  & $    0.993   $ & $    0.987      $ & $  0.909    $\\
 E1  & $    0.010  $ & $   1.0\times10^{-12}$ & $  0.0$\\
 E2  & $    0.990    $ & $   0.8248529      $ & $   7.25\times 10^{-3}$\\
 E3  & $    0.915      $ & $   0.0236           $ & $  1.71\times 10^{-14}$\\
 F1  & $    0.148      $ & $   1.15\times 10^{-9}   $ & $  0.0$\\
 F2  & $    0.401      $ & $   1.07\times 10^{-4}   $ & $  3.25\times 10^{-23}$\\
 G1  & $    0.001     $ & $  6.64\times 10^{-26}   $ & $  0.0$\\
 G2  & $ 1.42\times 10^{-17}$ & $  0.0          $ & $   0.0$\\
 H1  & $    0.996      $ & $  0.9941702        $ & $  0.990   $\\ 
 H2  & $    0.990     $ & $   0.9725315      $ & $  0.420  $\\  
 I  & $     0.989     $ & $   0.9746811       $ & $ 0.592   $\\ 
 J  & $     0.992     $ & $   0.9898050       $ & $ 0.953   $\\ 
 L1  & $    0.461     $ & $   5.59\times 10^{-5} $ & $  2.69\times 10^{-27}$\\
 L2  & $    0.188     $ & $   4.95\times 10^{-9}  $ & $ 0.0$\\
 M1  & $    0.995     $ & $   0.9549476           $ & $ 0.111  $\\  
 M2  & $    0.985     $ & $   0.9328853         $ & $ 0.405  $\\  
 N  & $     0.995     $ & $   0.9818770        $ & $  0.365  $\\  
 O  & $  6.16\times10^{-12}  $ & $   0.00      $ & $  0.0$\\
\end{tabular}

\end{center}
\label{wyn}
\end{table}

\section{CONSTRAINING STELLAR EVOLUTIONARY PARAMETERS}

In this section we discuss a way of quantifying the results presented
above. In order to test the usefulness of observations of chirp masses for
constraining stellar evolution models we performed the a Monte Carlo
simulation of such observations. We have assumed that the underlying
stellar population is described by a given model chosen from
Table~\ref{tone}.  We estimated if one can reject a hypothesis that the
stellar population is described by the standard model A with an
observation of 20, 100, and 500 mergers. We chose ten thousand random
realizations of the observed sample of mergers. Each such realization was
compared with the distribution of model A using the Kolmogorov Smirnov
(KS) test.  We then examined the distribution of the KS-test probabilities
and found the probability that appeared in less than 1\% of simulations. A
high value of this probability means that the given model can not be
distinguished from model A, while a low value means that the two models
can easily be distinguished. We present the results in Table~\ref{wyn}. We
first notice that models D do not differ from model A, just as expected.
In the case of models that affect the distribution of masses and
especially the range of their masses like e.g. models G and O, already a
small sample of several tens of mergers will be sufficient to recognize
them.  With a larger sample of a hundred mergers one should be able to
constrain the common envelope parameter $\alpha_{CE}\lambda$ (models E).
Also it should be possible to determine the parameters describing mass
transfer events: the mass fraction accreted in mass transfer events
(models F), and the angular momentum of the material lost in mass
transfers. However, a larger sample of mergers will be required to
constrain the kick velocity distribution: even with 500 detections it is
going to be possible to put only rough constraints on the width of the
kick velocity distribution.

\section{CONCLUSIONS}

Using the StarTrack population synthesis code we show that  the distribution of
masses of merging compact objects carries  useful information about the ways
massive stars evolve. In particular the stellar wind strength, common envelope
efficiency,  loss of angular momentum efficiency, and the fraction of mass 
accreted in mass transfer events leave distinct tracks on the  distribution of
observed chirp masses. The Monte Carlo simulations show that analysis of this 
distribution shall yield  useful  constraints on the evolutionary parameters of
binaries containing massive stars. The advanced LIGO configuration will  be
able to gather sufficient number of  merging events to perform such analysis.

\acknowledgments     
 This research was funded by KBN through grant number 5P03D01120. TB is
grateful for the hospitality of the Theoretical Astrophysics Group at
Northwestern University. KB acknowledges support by the Lindheimer Fund at
Northwestern University. VK acknowledges partial support by NSF grant
PHY-0121420.

\newcommand\apj{ApJ}\newcommand\mnras{MNRAS}\newcommand\apjl{ApJLett}
\newcommand\prd{Phys ReV D}

\begin{thebibliography}{10}

\bibitem{1992Sci...256..325A}
A.~{Abramovici}, W.~E. {Althouse}, R.~W.~P. {Drever}, Y.~{Gursel},
  S.~{Kawamura}, F.~J. {Raab}, D.~{Shoemaker}, L.~{Sievers}, R.~E. {Spero}, and
  K.~S. {Thorne}, ``{LIGO - The Laser Interferometer Gravitational-Wave
  Observatory},'' {\em Science} {\bf 256}, pp.~325--333, Apr. 1992.

\bibitem{1990brada}
Bradaschia {\em et~al.}, ``The virgo detector,'' {\em Nucl. Instrum and
  Methods} {\bf A289}, p.~518, 1990.

\bibitem{2001ApJ...556..340K}
V.~{Kalogera}, R.~{Narayan}, D.~N. {Spergel}, and J.~H. {Taylor}, ``{The
  Coalescence Rate of Double Neutron Star Systems},'' {\em \apj} {\bf 556},
  pp.~340--356, July 2001.

\bibitem{2002ApJ...572..407B}
K.~{Belczynski}, V.~{Kalogera}, and T.~{Bulik}, ``{A Comprehensive Study of
  Binary Compact Objects as Gravitational Wave Sources: Evolutionary Channels,
  Rates, and Physical Properties},'' {\em \apj} {\bf 572}, pp.~407--431, June
  2002.



\bibitem{KKL}
C.-L.\ Kim, V.\ Kalogera, D.R.\ Lorimer, ApJ, submitted, astro-ph/0207408
``The Probability Distribution of Binary Pulsar Coalescence Rates. I.
Double Neutron Star Systems in the Galactic Field''


\bibitem{will}
C.~M. Will, ``The confrontation between general relativity and experiment,''
  {\em Living Rev. Relativity} {\bf 4}, p.~4, 2001.
\newblock http://www.livingreviews.org/Articles/Volume4/2001-4will/.

\bibitem{blanchet}
L.~Blanchet, ``Gravitational radiation from post-newtonian sources ans
  inspiralling compact binaries,'' {\em Living Rev. Relativity} {\bf 5}, p.~3,
  2002.
\newblock http://www.livingreviews.org/Articles/Volume5/2002-3blanchet/.

\bibitem{1999MNRAS.308..153H}
W.~C.~G. {Ho} and D.~{Lai}, ``{Resonant tidal excitations of rotating neutron
  stars in coalescing binaries},'' {\em \mnras} {\bf 308}, pp.~153--166, Sept.
  1999.

\bibitem{2000MNRAS.315..543H}
J.~R. {Hurley}, O.~R. {Pols}, and C.~A. {Tout}, ``{Comprehensive analytic
  formulae for stellar evolution as a function of mass and metallicity},'' {\em
  \mnras} {\bf 315}, pp.~543--569, July 2000.

\bibitem{1992ApJ...391..246P}
P.~{Podsiadlowski}, P.~C. {Joss}, and J.~J.~L. {Hsu}, ``{Presupernova evolution
  in massive interacting binaries},'' {\em \apj} {\bf 391}, pp.~246--264, May
  1992.

\bibitem{1984ApJ...277..355W}
R.~F. {Webbink}, ``{Double white dwarfs as progenitors of R Coronae Borealis
  stars and Type I supernovae},'' {\em \apj} {\bf 277}, pp.~355--360, Feb.
  1984.

\bibitem{1999ApJ...522..413F}
C.~L. {Fryer}, ``{Mass Limits For Black Hole Formation},'' {\em \apj} {\bf
  522}, pp.~413--418, Sept. 1999.

\bibitem{1998ApJ...505..315C}
J.~M. {Cordes} and D.~F. {Chernoff}, ``{Neutron Star Population Dynamics. II.
  Three-dimensional Space Velocities of Young Pulsars},'' {\em \apj} {\bf 505},
  pp.~315--338, Sept. 1998.

\bibitem{2002ApJ...567L..63B}
K.~{Belczynski}, T.~{Bulik}, and W.~{\l}. {Klu{\' z}niak}, ``{Population
  Synthesis of Neutron Stars, Strange (Quark) Stars, and Black Holes},'' {\em
  \apjl} {\bf 567}, pp.~L63--L66, Mar. 2002.

\bibitem{1993PhRvD..47.2198F}
L.~S. {Finn} and D.~F. {Chernoff}, ``{Observing binary inspiral in
  gravitational radiation: One interferometer},'' {\em \prd} {\bf 47},
  pp.~2198--2219, Mar. 1993.

\bibitem{1998PhRvD..57.4535F}
E.~E. {Flanagan} and S.~A. {Hughes}, ``{Measuring gravitational waves from
  binary black hole coalescences. I. Signal to noise for inspiral, merger, and
  ringdown.},'' {\em \prd} {\bf 57}, pp.~4535--4565, 1998.




\end{thebibliography}

\bibliographystyle{spiebib}   

\end{document}